\documentclass[12pt]{JHEP}
\usepackage{epsfig}
\usepackage{equations}
\newenvironment{Eqnarray}%
     {\arraycolsep 0.14em\begin{eqnarray}}{\end{eqnarray}}

\def\beq{\begin{equation}}
\def\eeq{\end{equation}}
\def\beqa{\begin{Eqnarray}}

\def\eeqa{\end{Eqnarray}}
\def\tanb{\tan\beta}
\def\sinb{\sin\beta}
\def\cosb{\cos\beta}

\def\sinbma{\sin(\beta-\alpha)}
\def\cosbma{\cos(\beta-\alpha)}

\def\hl{h}
\def\ha{A}
\def\hh{H}
\def\hpm{H^\pm}
\def\mha{m_{\ha}}
\def\mhl{m_{\hl}}
\def\mhh{m_{\hh}}
\def\mhpm{m_{\hpm}}

\def\mz{m_Z}
\def\mw{m_W}

\def\amu{\delta a_\mu^{\rm NP}}
\def\ifmath#1{\relax\ifmmode #1\else $#1$\fi}
\def\ls#1{\ifmath{_{\lower1.5pt\hbox{$\scriptstyle #1$}}}}
\def\half{\ifmath{{\textstyle{1 \over 2}}}}
\def\eq#1{eq.~(\ref{#1})}

\def\Ref#1{ref.~\cite{#1}}
\def\Refs#1#2{refs.~\cite{#1} and \cite{#2}}

\def\phm{\phantom{-}}
\def\lsim{\mathrel{\raise.3ex\hbox{$<$\kern-.75em\lower1ex\hbox{$\sim$}}}}
\def\gsim{\mathrel{\raise.3ex\hbox{$>$\kern-.75em\lower1ex\hbox{$\sim$}}}}
\newcommand{\mathbold}[1]{\mbox{\boldmath $\bf#1$}}

\title{Can the Higgs sector contribute significantly to the
muon anomalous magnetic moment?}

\author{Athanasios Dedes \\ 
Physikalisches Institut der Universit\"at Bonn,\\ 
 Nu\ss allee 12, D-53115 Bonn, Germany\\
E-mail: \email{dedes@th.physik.uni-bonn.de}}

\author{Howard E. Haber  \\ 
Santa Cruz Institute for Particle Physics\\ 
University of California, Santa Cruz, CA, 95064 USA\\
E-mail: \email{haber@scipp.ucsc.edu}}

\abstract{A light CP-even Higgs boson with $\mhl\sim 10$~GeV could
explain the recent BNL measurement of the muon anomalous magnetic
moment, in the framework of a general CP-conserving
two-Higgs-doublet extension of
the Standard Model with no tree-level flavor-changing neutral Higgs 
couplings.  However, the allowed Higgs mass window is quite
small and the corresponding model parameters are very constrained.
The Higgs sector can contribute significantly to the observed BNL
result for $g-2$ without violating known experimental constraints only
if the $\hl ZZ$ coupling (approximately) vanishes and
$M_{\Upsilon}\lsim\mhl\lsim 2m_B$.}


\preprint{SCIPP-01/10 \\ hep-ph/0102297}

\begin{document}

\section{Introduction}

A new experimental value of the muon anomalous
magnetic moment, $a_\mu\equiv\half(g-2)_\mu$, measured at BNL,
was recently reported in \Ref{Brown:2001mg}.
Comparing the measured value to its predicted value in the
Standard Model (SM), \Ref{Brown:2001mg} reported that
\begin{eqnarray}
a_\mu^{\rm exp}-a_\mu^{\rm SM}=426\pm 165 \times 10^{-11}\;.
\label{dev}
\end{eqnarray}
Ref.~\cite{Czarnecki:2001pv} has reviewed the Standard Model
computation of $a_\mu$ and concluded that if the deviation
of \eq{dev} can be attributed to new physics effects [$\amu$],
then at $90\%$~CL, $\amu$ must
lie in the range
\begin{eqnarray}
215 \times 10^{-11} \lsim \amu \lsim 637
 \times 10^{-11} \;.
\label{newphys}
\end{eqnarray}
This contribution is positive, and is of the order  
of the electroweak corrections to $a_\mu$.
More precisely, the contribution needed from new physics effects has to be 
of the order of $G_Fm_\mu^2/(4\pi^2\sqrt{2})$, where
$G_F=1.16637(1)\times 10^{-5}~{\rm GeV}^{-2}$ is Fermi's constant.
In this paper, we consider the possibility that $\amu$
arises entirely from the Higgs sector.  In the SM,
the Higgs boson contribution to $a_\mu$ is further suppressed
(relative to the main electroweak contribution) by a factor of
$m_\mu^2/\mhl^2$.  In light of the recent SM Higgs mass limit,
$\mhl\gsim 113.5$~GeV obtained at the LEP collider \cite{higgslimit},
the SM Higgs contribution to $a_\mu$ is clearly negligible.  

However, the Higgs sector contribution to $a_\mu$ could be
considerably enhanced in a two-Higgs-doublet extension of the Standard
Model (2HDM).  The significance of the $(g-2)_\mu$ constraint for the
2HDM (in light of the LEP Higgs constraints) was emphasized in
\Ref{Krawczyk:1997sm}, where the constraints of the previous BNL $(g-2)_\mu$
measurements were analyzed and the implications of future $(g-2)_\mu$
measurements were considered.\footnote{In \Ref{Krawczyk:1997sm},
it was assumed that the Higgs--fermion interaction was not the most
general, but of a form that guarantees the absence of tree-level
flavor--changing neutral Higgs couplings.  Alternatively,
one could assume the most general Higgs--fermion interaction
(thereby generating tree-level Higgs-mediated 
flavor--changing neutral currents [FCNCs]), and
choose the parameters of the model to avoid conflict with experimental
limits on FCNCs.  For example, such a model would possess a tree-level
$\hl \mu^\pm\tau^\mp$ coupling, which could contribute significantly
to $(g-2)_\mu$ \cite{sher}.  We choose not to consider a 2HDM with
flavor-changing neutral Higgs couplings in this paper.}  
Now that we have the first possible
indication of $\amu\neq 0$, it is appropriate to revisit the question
of the Higgs sector contribution to $a_\mu$.

The enhancement of the Higgs sector contribution to $a_\mu$ relative to
the SM result can arise from two different effects.
First, an enhanced $\hl\mu^+\mu^-$ coupling proportional to the ratio of
Higgs vacuum expectation values, $\tan\beta$, yields a Higgs
contribution to $\amu$ proportional to $\tan^2\beta$.  Second, a
suppressed $\hl ZZ$ coupling, proportional to $\sin(\beta-\alpha$)
[using notation reviewed below], can permit the existence of a CP-even
Higgs boson mass substantially below the LEP SM Higgs mass
limit.  In units of $G_Fm_\mu^2/(4\pi^2\sqrt{2})$,
the overall enhancement is of order  
\begin{equation}
\frac{m_\mu^2}{\mhl^2}\times \tan^2\beta \times
F\left(\frac{m_\mu^2}{m_h^2}\right)\simeq 1{\hbox{\rm---}}10 \;.
\end{equation}
$F(x)$ is a loop factor which involves logarithms 
of the form $\ln(\mhl^2/m_\mu^2) \sim {\cal O}(10)$.
A light CP-even Higgs boson with $\mhl \simeq 10$ GeV
and $30 \lsim \tan\beta \lsim 50$, predicts a muon anomalous magnetic
moment to lie in the $90\%$~CL allowed range for new physics effects
specified in \eq{newphys}.

A 2HDM in which the Higgs sector contribution to $\amu$ is significant
is not compatible with the Higgs
sector of the minimal supersymmetric extension of the Standard Model
(MSSM).  This is true because one cannot have a very light $\hl$ with
suppressed $\hl ZZ$ couplings without an observable rate for
$Z\to \hl\ha$, in conflict with LEP 
data \cite{susyhiggs}.
Moreover, the MSSM provides additional mechanisms for generating
significant contributions to $\amu$.  A number of recent 
papers \cite{Everett:2001tq,Feng:2001tr,
Baltz:2001ts,Chattopadhyay:2001vx,Komine:2001fz} have
shown that the recent BNL measurement is compatible with
supersymmetric contributions to $\amu$ involving chargino and
neutralino exchange, over an interesting region of MSSM parameter space.



In this paper, we focus on the possibility that the new physics
contribution to $a_\mu$ arises solely from the Higgs sector.
The two-doublet Higgs sector~\cite{Gunion:1989we} contains eight scalar
degrees of freedom.  It is convenient to distinguish between the
two doublets by employing 
one complex $Y=-1$ doublet, {\boldmath $\Phi_d$}$=(\Phi_d^0,\Phi_d^-)$
and one complex $Y=+1$ doublet, {\boldmath
$\Phi_u$}$=(\Phi_u^+,\Phi_u^0)$.  To avoid tree-level Higgs-mediated flavor
changing neutral currents, we do not allow the most general
Higgs--fermion interaction \cite{glashwein}.
Instead, we impose discrete symmetries
(which may be softly-broken by mass terms), and consider two possible
models \cite{hallwise}.  
In Model~I, $\Phi_d^0$ couples to both up-type and
down-type quark and lepton pairs, while the coupling of $\Phi_u^0$ to
fermion pairs is absent.\footnote{One can just as well assume that
$\Phi_u^0$ couples to both up-type and
down-type quark and lepton pairs, while the coupling of $\Phi_d^0$ to
fermion pairs is absent.  In this case, all the results of this
paper would apply simply by replacing $\tan\beta$ with $\cot\beta$.}  
In Model~II, $\Phi_d^0$
[$\Phi_u^0$] couples exclusively to down-type [up-type] fermion pairs.
When the Higgs potential is minimized, the neutral components of the
Higgs fields acquire vacuum expectation values:\footnote{In this
paper, we neglect the possibility of significant CP-violation in the
Higgs sector.  In this case, the
phases of the Higgs fields can be chosen such that the vacuum
expectation values are real and positive.}
\beq
\langle {\mathbold{\Phi_d}} \rangle={1\over\sqrt{2}} \left(
\begin{array}{c} v_d\\ 0\end{array}\right), \qquad \langle
{\mathbold{\Phi_u}}\rangle=
{1\over\sqrt{2}}\left(\begin{array}{c}0\\ v_u

\end{array}\right)\,,\label{potmin}
\eeq
where the normalization has been chosen such that
$v^2\equiv v_d^2+v_u^2=(246~{\rm GeV})^2$,
while the ratio $\tan\beta\equiv v_u/v_d$ is a free parameter of the
model.  The physical Higgs spectrum consists of
a charged Higgs pair
\beq \label{hpmstate}
\hpm=\Phi_d^\pm\sinb+ \Phi_u^\pm\cosb\,,
\eeq
one CP-odd scalar
\beq \label{hastate}
\ha= \sqrt{2}\left({\rm Im\,}\Phi_d^0\sinb+{\rm Im\,}\Phi_u^0\cosb
\right)\,,
\eeq
and two CP-even scalars:
\beqa
\hl &=& -(\sqrt{2}\,{\rm Re\,}\Phi_d^0-v_d)\sin\alpha+
(\sqrt{2}\,{\rm Re\,}\Phi_u^0-v_u)\cos\alpha\,,\nonumber\\
\hh &=& (\sqrt{2}\,{\rm Re\,}\Phi_d^0-v_d)\cos\alpha+
(\sqrt{2}\,{\rm Re\,}\Phi_u^0-v_u)\sin\alpha\,,
\label{scalareigenstates}
\eeqa
(with $\mhl\leq \mhh$).
The angle $\alpha$ arises when the CP-even Higgs
squared-mass matrix (in the $\Phi_d^0$---$\Phi_u^0$ basis) is
diagonalized to obtain the physical CP-even Higgs states.

We briefly review the Higgs couplings relevant for our
analysis.  The tree-level $\hl$ couplings to $ZZ$ and $\ha Z$ are given by
\beqa
g\ls{\hl ZZ} &=& {g\mz\sinbma\over\cos\theta_W}\,,\label{vvcoup}\\
g\ls{\hl\ha Z} &=& {g\cosbma\over 2\cos\theta_W}\,.
           \label{hvcoup}
\eeqa
For the corresponding couplings of $\hh$ to $ZZ$ and $\ha Z$,
one must interchange $\sinbma$ and $\cosbma$ in the above formulae.

The pattern of couplings of the Higgs bosons to fermions depends 
on the choice of model.  However, in this paper we are mainly concerned
with the coupling of down-type fermions to Higgs bosons, which are the
same in Model~I and Model~II.  
For our analysis, the relevant 
couplings of the neutral Higgs bosons to $b\bar b$ or $\mu^+\mu^-$
relative to the SM
value, $m_f/v$ [$f$= $b$ or $\mu$], are given by
\begin{eqaligntwo}
 \label{qqcouplings}
\hl b\bar b \;\;\; ({\rm or}~ \hl \mu^+ \mu^-):&~~~ -
{\sin\alpha\over\cos\beta}=\sin(\beta-\alpha)
-\tan\beta\cos(\beta-\alpha)\,,\\[3pt]
\hh b\bar b \;\;\; ({\rm or}~ \hh \mu^+ \mu^-):&~~~
\phm{\cos\alpha\over\cos\beta}=
\cos(\beta-\alpha)
+\tan\beta\sin(\beta-\alpha)\,,\\[3pt]
\ha b \bar b \;\;\; ({\rm or}~ \ha \mu^+
\mu^-):&~~~\phm\gamma_5\,{\tan\beta}\,,
\end{eqaligntwo}
(the $\gamma_5$ indicates a pseudoscalar coupling), and the
charged Higgs boson couplings to muon pairs
(with all particles pointing into the vertex) is given by
\beq
\label{hpmqq}
g_{H^- \mu^+ \nu}= {gm_\mu\over{\sqrt{2}\mw}}\
\tan\beta\,P_L\,,
\eeq
where $P_L\equiv\half(1-\gamma_5)$.

We have noted above that only light Higgs bosons with enhanced
couplings to down-type fermions can contribute appreciably to $\amu$.
To avoid the LEP SM Higgs mass limit, such a light Higgs boson should be
almost decoupled from $ZZ$.  This implies that either $\hl$ is light,
with $|\sin(\beta-\alpha)|\ll 1$ [see \eq{vvcoup}] or $\ha$ is light
(since $\ha$ has no tree-level coupling to vector boson pairs).  In
the next section, we will show that a light $\ha$ makes a {\it
negative} contribution to $\amu$ and thus is not compatible with the
recent BNL measurement.  Hence, we focus on the 2HDM in which only
$\hl$ is light and $\sin(\beta-\alpha)\simeq 0$.  From
\eq{qqcouplings}, we see that if $\sin(\beta-\alpha)\simeq 0$, then
the coupling of $\hl$ to down-type fermions is proportional to
$\tan\beta$.  Thus, in the region of large $\tan\beta$ and small
$\sin(\beta-\alpha)$, the contribution of a light CP-even Higgs boson
of the 2HDM may yield a significant correction to $\amu$ without
being in conflict with the LEP SM Higgs search.

Although the considerations above apply to both Model~I and Model~II,
it is important to note that the Higgs couplings to up-type fermions
differ between the two models.  The Model~II
$\hl t\bar t$ coupling relative to its SM value, $m_t/v$,
is given by:
\beq
 \label{ttcouplings}
\hl t\bar t~:~~~ 
{\cos\alpha\over\sin\beta}=\sin(\beta-\alpha)
+\cot\beta\cos(\beta-\alpha)\,,
\eeq
whereas the Model~I $\hl t\bar t$ coupling relative to $m_t/v$
is the same as the Model~II $\hl b\bar b$ coupling
relavive to $m_b/v$.
That is, for $\sinbma=0$, the Model~II 
$\hl t \bar t$ coupling is proportional to $\cot\beta$ and is
therefore suppressed at large $\tanb$, while in
Model~I, $|g_{\hl t\bar t}|=(m_t/v)\tanb\gg 1$.
Thus, the $\tanb$ enhanced Model~I Higgs couplings to $t\bar t$ are
non-perturbative at large $\tanb$.  
Both theoretical and experimental considerations
lead us to reject this possibility.  Henceforth, we will assume that
the 2HDM contains Model~II Higgs--fermion couplings.

Finally, we note that in the parameter region cited above, the heavier
Higgs bosons, $\hh$, $\ha$, $\hpm$, cannot be arbitrarily heavy.  If
one attempts to take such a limit, one finds that there must be some
Higgs quartic self-couplings that become significantly larger 
than 1 \cite{decoupling}.
That is, this model does not possess a decoupling limit.  However,
the model stays weakly coupled as long as the heavier Higgs states are
not too much larger than $v=246$~GeV.   In contrast, in the limit of
$\cos(\beta-\alpha)=0$, the couplings of $\hl$ reduce to those of the
SM Higgs boson.  This decoupling limit can be formally
reached by taking the masses of $\hh$, $\ha$, $\hpm$ to be arbitrarily
large, while keeping the quartic Higgs self-couplings 
$\lsim {\cal O}(1)$ \cite{decoupling}.  
The resulting low-energy effective theory is just the
SM with one Higgs doublet.  Of course, as we have noted above, the
contribution of SM Higgs boson to $\amu$ is negligible.
Thus, over an intermediate range of heavy Higgs masses, the 
contributions of
$\hh$, $\ha$, $\hpm$ (which are $\tan^2\beta$ enhanced) to $\amu$
will be significantly larger than that of $\hl$ even though
$\cosbma\simeq 0$.

\section{Model~II Higgs boson corrections to the muon anomalous magnetic
  moment}

The first calculation of the one-loop electroweak corrections to the muon
anomalous magnetic moment was presented by Weinberg and
Jackiw~\cite{Jackiw:1972jz} and 
by Fujikawa, Lee and Sanda~\cite{Fujikawa:1972fe}.  A very useful
compendium of formulae for the one-loop corrections to $g-2$
in a general electroweak model was
given in \Ref{Leveille:1978rc}, 
and applied to the 2HDM in \Ref{haber}.\footnote{Here, we correct
a small error in the expression in the $\hpm$ contribution
given in \Ref{haber}.} 
In the 2HDM, both neutral and charged Higgs bosons contribute to $g-2$.
A convenient list of the relevant formulae can be found in
\Ref{Krawczyk:1997sm}.

\begin{eqnarray}
\delta a_\mu^h &=& \frac{G_Fm_\mu^2}{4\pi^2 \sqrt{2}}~
\biggl (\frac{\sin{\alpha}}{\cos{\beta}}\biggr )^2 ~R_h~ F_h(R_h) \; \\
\delta a_\mu^H &=& \frac{G_Fm_\mu^2}{4\pi^2 \sqrt{2}}~
\biggl (\frac{\cos{\alpha}}{\cos{\beta}}\biggr )^2 ~R_H~ F_H(R_H) \; \\
\delta a_\mu^A &=& \frac{G_Fm_\mu^2}{4 \pi^2 \sqrt{2}}~
\tan^2\beta ~R_A~ F_A(R_A) \; \\
\delta a_\mu^{H^\pm} &=& \frac{G_Fm_\mu^2}{4 \pi^2 \sqrt{2}}~
\tan^2\beta ~R_{H^\pm}~ F_{H^\pm}(R_{H^\pm},R_\nu) \;
\label{amuch}
\end{eqnarray}
where $R_{h,H,A,H^\pm}\equiv m_\mu^2/m_{h,H,A,H^\pm}^2$,
$R_\nu\equiv m_\nu^2/m_{H^\pm}^2$ and 
\begin{eqnarray}
F_{h,H}(R_{h,H}) &=& \int_0^1 dx\, \frac{x^2(2-x)}{R_{h,H} x^2-x+1}
 \;, \\
F_{A}(R_{A}) &=& \int_0^1 dx\, \frac{-x^3}{R_{A} x^2-x+1}
 \;, \\
F_{H^\pm}(R_{H^\pm},R_\nu) &=& \int_0^1 dx\, \frac{-x^2(1-x)}{R_{H^\pm}
 x^2+(1-R_{H^\pm}-R_\nu) x+ R_\nu}
 \;, 
\end{eqnarray}
The neutrino mass is negligible, so henceforth we set $R_\nu=0$.
Since $R_{h,H,A,H^\pm}\ll 1$, one can easily expand the above
integrals in the corresponding small parameter.  In the next two
subsections, we write out the leading terms in this expansion, which
are quite accurate in the Higgs mass range of interest.\footnote{The
plot shown in this paper is based on the exact values of the above
integrals.}   

\subsection{Non-decoupling limit: $\sin(\beta-\alpha)=0$}

In section 1, we argued that the most significant Higgs contribution
to $\amu$ (consistent with the LEP SM Higgs search) arises in the
parameter regime in which $\sinbma\simeq 0$ and $\tanb\gg 1$.  
Setting  $\sinbma=0$ and keeping only the leading terms in 
$R$ when evaluating the above integrals, 
the total Higgs sector contribution to $a_\mu$
is given by:
\begin{eqnarray}
&& \delta a_\mu^{\rm Higgs}=  \delta a_\mu^h+\delta a_\mu^H
+\delta a_\mu^A+
\delta a_\mu^{H^\pm}  \nonumber \\[5pt]
&& \qquad\simeq\frac{G_Fm_\mu^2}{4\pi^2 \sqrt{2}}
\tan^2\beta \bigg \{
\frac{m_\mu^2}{\mhl^2} \biggl[\ln\left(\frac{\mhl^2}{m_\mu^2}\right)
-\frac{7}{6} \biggr] -\frac{m_\mu^2}{\mha^2} \biggl[
\ln\left(\frac{\mha^2}{m_\mu^2}\right) -\frac{11}{6}\biggr]-
\frac{m_\mu^2}{6m_{H^{\pm}}^2} \biggr \} \;. \nonumber \\
\label{hfull}
\end{eqnarray}
Note that the 
logarithms appearing in \eq{hfull} always dominate the corresponding
constant terms when the Higgs masses are larger than 1 GeV.
It is then clear that $\ha$ and $\hpm$ exchange contribute a
negative value to $\amu$.  Since our goal is to explain the BNL $g-2$
measurement which suggests a positive value for $\amu$, we should take
$\mha$ and $\mhpm$ large (masses above 100~GeV are sufficient)
in order that the corresponding
$\ha$ and $\hpm$ negative contributions are 
neglibly small.\footnote{Grifols 
and Pascual~\cite{Grifols:1980yk}
found that for a very light charged Higgs boson, 
the two-loop contribution to $a_\mu$ is positive and
can be larger in magnitude than the one-loop
result given in \eq{amuch}:
\begin{eqnarray} 
\delta
 a_\mu^{H^\pm}=a_\mu^{H^\pm}\hbox{\rm {(1--loop)}}+
\frac{1}{180}\left(\frac{\alpha}{\pi}\right)^2
\left(\frac{m_\mu}{m_{H^\pm}}\right)^2+{\cal
 O}\biggl [\left(\frac{m_\mu}{m_{H^\pm}}\right)^4 
\ln\left(\frac {m_\mu}{m_{H^\pm}}\right)
 \biggr ]  \;.
\end{eqnarray}
However, the LEP bound
on the charged Higgs mass \Ref{tomjunk},
$\mhpm>78.7$~GeV, implies that both the one and two-loop charged Higgs
contribution to $\amu$ are negligible.}
If $\amu$ is to be a consequence of the Higgs sector, it must be entirely
due to the contribution of the light CP-even Higgs boson.
Note that the heavier
CP-even Higgs, $\hh$, does not give a contribution proportional to
$\tan\beta$ (as shown in section 1); hence its contribution
to $\amu$ can be neglected in \eq{hfull}.
Thus, to a good approximation,
   
\FIGURE[t]{\centerline{\epsfig{figure=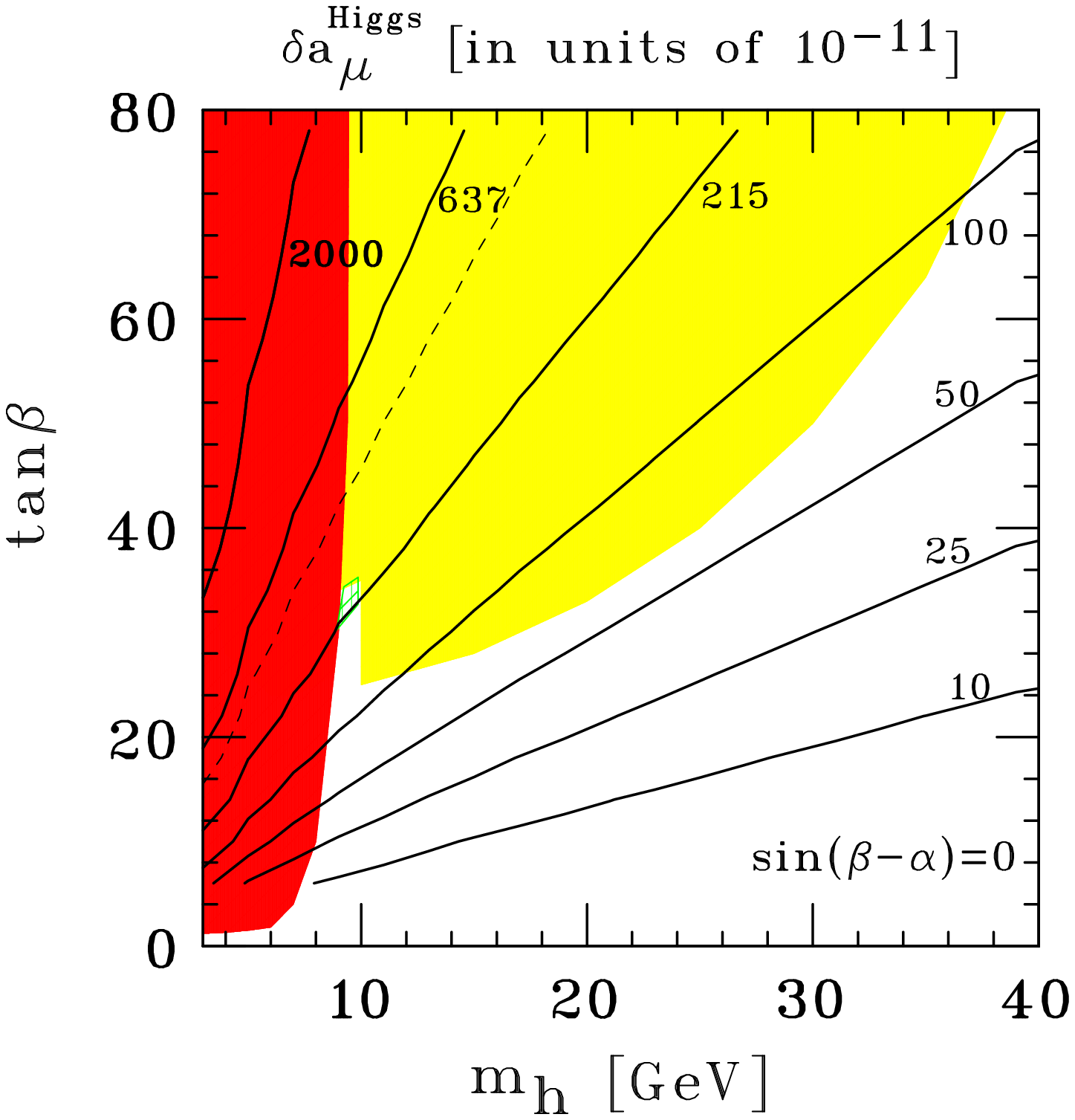,height=4in,angle=90}}
\caption{Contours of the  predicted one-loop Higgs sector contribution
to the muon anomalous magnetic moment, 
$\delta \alpha_\mu^{\rm Higgs}$ (in units of $10^{-11}$)
in the 2HDM, assuming that $\sin(\beta-\alpha)=0$, 
and $\mhh=\mha=\mhpm=200$~GeV
(there is little sensitivity to the heavier Higgs masses).
The dashed line contour corresponds to the central value of
$\delta a_\mu\equiv a_\mu^{\rm exp}-a_\mu^{\rm SM}$, as reported 
in \Ref{Brown:2001mg}.
The contour lines marked 215 and 637 correspond to 90\%~CL limits
for the contribution of new physics to $a_\mu$ [\eq{newphys}].
The dark-shaded (red) region is excluded 
by the CUSB Collaboration search for $\Upsilon\to\hl\gamma$ 
at CESR~\cite{CUSB}.  The light-shaded (yellow) region 
is excluded at $95\%$~CL by the ALEPH and DELPHI searches
for $e^+e^-\to \hl f\bar f$ ($f=b$ or $\tau$) at LEP~\cite{Aleph,Delphi}.
In the small hatched region (green) nestled between the 
two experimentally excluded
shaded regions, above the 215 contour line and 
centered around $\mhl\simeq 10$~GeV, the Higgs sector
contribution to $\amu$ lies within the $90\%$~CL allowed range
[\eq{newphys}].}}

\beq
\delta a_{\mu}^{\rm Higgs}\simeq 
\delta a_\mu^{{\hl}} \simeq \frac{G_Fm_\mu^2}{4\pi^2 \sqrt{2}}
\left(\frac{m_\mu^2}{\mhl^2}\right) \tan^2\beta
\bigg[
\ln\left(\frac{\mhl^2}{m_\mu^2}\right)-\frac{7}{6} \biggr ] \;.
\label{app1}
\eeq

One can check that a light Higgs boson with a mass of
around 10 GeV and with $\tan\beta=35$ gives $\delta a_{\mu}^{\rm
Higgs}\simeq 280 \times 10^{-11}$, which is within the
90\%~CL allowed range for $\amu$ quoted in \eq{newphys}.
Contour lines corresponding to a full numerical 
evaluation of the Higgs sector one-loop 
contribution to $\delta
a_\mu^{\rm Higgs}$ [in units of $10^{-11}$] are exhibited
in fig.~1, for $\sinbma=0$ and $\mhh=\mha=\mhpm=200$~GeV.\footnote{%
The results are insensitive to the values of the heavy Higgs masses
above 100~GeV.}
The relevant experimental bounds are also displayed
in fig.~1; these limits are reviewed in section 3.
A careful inspection of the excluded region in the $\mhl$ {\it vs.}
$\tanb$ parameter space shows that a light
Higgs boson of around 10 GeV mass and $30\lsim \tan\beta \lsim 35$
is permitted.  In this parameter regime, we obtain
a value for $\amu$ within the $90\%$~CL allowed range of
\eq{newphys}.  However, the central value of $\amu$ given
in \eq{newphys} lies within the excluded regions of fig.~1.

\subsection{Decoupling limit: $\cos(\beta-\alpha)=0$}

In the decoupling limit, where $\cosbma\simeq 0$ and $\mha\gg\mz$,
the couplings of the light Higgs boson,
$\hl$, are (nearly) identical to those of the SM Higgs boson.
As a result, the LEP SM Higgs mass bound of $\mhl\gsim 113.5$~GeV applies.
For $\cosbma=0$, the $\hh$ couplings to down-type fermion pairs are 
enhanced by $\tanb$ [see \eq{qqcouplings}].
Thus, the Higgs sector contribution to $\amu$ is given by
\eq{hfull}, with $\mhl$ replaced by $\mhh$.  In the decoupling limit,
$\mhh\simeq\mha\simeq \mhpm$ [the mass differences are of ${\cal
O}(\mz^2/\mha)$].  Setting $\cosbma=0$ and $\mhh=\mha=\mhpm$, we find
\beq
\delta a_\mu^{\rm Higgs} \simeq  \frac{G_Fm_\mu^2}{4 \pi^2\sqrt{2}}
\left(\frac{m_\mu^2}{\mha^2}\right)\tan^2\beta \biggl[
\frac{1}{2}-\left(\frac{2m_\mu^2}{\mha^2}\right) 
\ln\left(\frac{\mha^2}{m_\mu^2}\right)
 \biggr ]\;.
\label{app2}
\eeq
The contribution of $\hl$ is not $\tanb$--enhanced and is thus
negligible.  In is interesting to note that for values of
$\mha\lsim\mhl\tanb$, the heavier (``decoupled'')
Higgs bosons actually dominate in the Higgs sector contribution
to $\amu$.\footnote{If we formally take $\mha\to\infty$, we recover
the Standard Model Higgs contribution to $a_\mu$.}
However, for 100~GeV $<\mha< 1000$~GeV, and $30<\tanb<100$,
the Higgs sector contribution to $a_\mu$ ranges from about $5\times
10^{-12}$ to $5\times 10^{-14}$, which is
three to five orders of magnitude below what
is needed to explain the BNL measurement of $a_\mu$.

%


\section{CESR and LEP constraints on a light Higgs boson}

Let us consider the 2HDM in which $\sinbma=0$, $\tanb\gg 1$
and $\mhl\sim {\cal O}(10$~GeV), which are necessary conditions
if the Higgs sector is to be the source for $\amu$ in the range
given by \eq{newphys}.
The $\hl\ha Z$ coupling is maximal [\eq{hvcoup}], so we 
must assume that $\mha$ is large enough so that
$e^+e^-\to \hl\ha$ is not observed at LEP.
The tree-level $\hl ZZ$ coupling is absent, which implies that
the LEP SM Higgs search based on $e^+e^-\to Z\to Z\hl$ does not impose
any significant constraints on $\mhl$.\footnote{Presumably,
radiative corrections would lead to a small effective value for
$\sinbma$.  The LEP Higgs search yields an excluded region in the 
$\sinbma$ {\it vs.} $\mhl$ plane, and implies that for $\mhl\sim 10$~GeV, 
$|\sinbma|\lsim 0.06$ is not excluded
at 95\%~CL~\cite{Janot,Sopczak:1995zy}.}  
However, there are a number of constraints on light Higgs masses 
that do not rely on the $\hl ZZ$ coupling.  For Higgs bosons with
$\mhl\lsim 5$~GeV, the SM Higgs boson was ruled out by a
variety of arguments that were summarized in \Ref{Gunion:1989we}.  For
5~GeV $\lsim\mhl\lsim 10$~GeV, the relevant
Higgs boson constraint can be derived 
from the absence of Higgs production in $\Upsilon\to\hl\gamma$.

An experimental search for $\Upsilon\to\hl\gamma$ by the
CUSB Collaboration at CESR~\cite{CUSB} 
found no candidates.  The Higgs mass
limit obtained from this result depends on the theoretical
prediction.  In addition to the non-relativistic,
tree-level prediction of \Ref{wil},
there are three classes of corrections that have been explored in the
literature: ${\cal O}(\alpha_s)$ hard QCD
corrections~\cite{vys,nason}, relativistic corrections to the
non-relativistic treatment of the $b\bar b$ bound state
\cite{biswas,azn}, and bound state threshold corrections \cite{wu}.
The theoretical picture that emerges is uncertain.  The hard QCD
corrections are large and suggest that ${\cal O}(\alpha_s^2)$
corrections could be significant.  In addition, relativistic effects
enter at the same order as the ${\cal O}(\alpha_s)$ corrections; both
are of ${\cal O}(v^2/c^2)$ and the two must be treated
consistently.  Finally,  \Ref{wu} argued that
strong cancellations can occur among various contributions in the
threshold region, leading to an additional suppression in rate of 
about 14 for $\mhl=8.5$~GeV (and even a larger suppression
as $\mhl\to M_{\Upsilon}$).  The application of the theoretical
analysis of $\Gamma(\Upsilon\to\hl\gamma)$ to the CUSB data suggests
that values of $\mhl\lsim 5$---7~GeV can be ruled out
at $95\%$~CL, although a
precise upper limit cannot be obtained due to 
the theoretical uncertainties outlined above.

The above discussion was relevant for obtaining a limit on the mass of
the SM Higgs boson.  In the 2HDM considered here, $\tanb\gg 1$, and
the prediction for $\Gamma(\Upsilon\to\hl\gamma)$ is enhanced by a
factor of $\tan^2\beta$.  For values of $\tanb\gsim 10$, the CUSB data
can reliably rule out Higgs masses up to about 8~GeV.  As $\mhl\to
M_\Upsilon$, the precise experimental limit is not very well known
due to the theoretical uncertainties 
near threshold mentioned above.  Our estimate
for the excluded region for $\mhl\lsim M_\Upsilon$ is
indicated by the dark (red) shaded region in fig.~1.  Note that
for Higgs masses above 8~GeV, $\tanb\gsim 30$ if the Higgs
sector contribution to $\amu$ lies in the $90\%$~CL range specified in
\eq{newphys}.  For such large values of $\tanb$, the predicted rate
for $\Upsilon\to\hl\gamma$ is increased by at least three orders of
magnitude relative to the SM.  This factor should dwarf the theoretical
uncertainties discussed above except for values of $\mhl$ very close
to $M_\Upsilon$.  Thus, in the 2HDM parameter regime of interest,
we obtain a lower bound of $\mhl\gsim M_\Upsilon$.

A second bound on $\mhl$ can be derived from the non-observation of 
Higgs bosons at LEP via the process $e^+e^-\to \hl f\bar f$ ($f=b$, $\tau$).
The cross-section for this process depends on the $\hl$ Yukawa
couplings to down-type fermions.
In the 2HDM with $\sin(\beta-\alpha)=0$, these Yukawa couplings are
enhanced (relative to the corresponding SM value) by 
$\tanb$.  Preliminary analyses 
by the ALEPH and DELPHI Collaborations at LEP
based on the search for $e^+e^-\to \hl f\bar f$ ($f=b$, $\tau$),
where $\hl\to
\tau^+\tau^-$, $b\bar b$, find no evidence for 
light Higgs boson production~\cite{Aleph,Delphi}.
Combining the two analyses, we
exclude at 95\%~CL the light-shaded (yellow) region of fig.~1.
Note that the lower limit on $\tanb$ changes discontinuously at 
$2m_B$, where $B$ is the lightest $B$-meson [$m_B=5.279$~GeV].  
For Higgs masses that lie in the range $2m_\tau\lsim\mhl\lsim 2m_B$,
the dominant Higgs decay mode is $\hl\to\tau^+\tau^-$.\footnote{By
assumption, $\tanb\gg 1$ and the rate for $\hl\to c\bar c$ is
suppressed by a factor of $\cot^2\beta$.}  In this mass range, the
ALEPH limit on $\tan\beta$ is better than the corresponding DELPHI
limit.  In
particular, for $M_{\Upsilon}\lsim\mhl\lsim 2m_B$, the ALEPH excluded
region implies that $\tanb\lsim 35$.  For values of
$\mhl>2m_B$, the Higgs decays primarily into $b\bar b$, and the DELPHI
limit (which is more powerful than the ALEPH limit in this mass range)
completely excludes the region of parameter space in which the
Higgs sector contribution to $\amu$ lies in the $90\%$~CL range 
specified in \eq{newphys}.

One other light Higgs process observable at LEP that is sensitive to
the Higgs--fermion Yukawa couplings, even in the absence of the $ZZ\hl$
and $W^+W^-\hl$
couplings, is the one-loop process $Z\to\hl\gamma$.  Both up-type and
down-type fermions contribute in the loop, so the decay rate 
in Model~I and Model~II differs.  Ref.~\cite{Krawczyk:1999kk} analyzes
the implication of this process for the general 2HDM with Model~II
couplings and shows that the LEP experimental constraints in the
$\mhl$ {\it vs.}  $\tanb$ plane for $\tanb>1$ are weaker than the ones
obtained from $e^+e^-\to\hl f\bar f$ discussed above.  In Model~I, we
can can use the results of \Ref{Krawczyk:1999kk} simply by
interchanging $\tanb$ and $\cot\beta$.  For $\mhl\sim 10$~GeV, the LEP
experimental constraints imply that $\tanb<10$.  Thus, we have an
independent reason to conclude that the Model~I 2HDM cannot provide an
explanation for the BNL measurement of $a_\mu$.

Finally, one must check the implications of the precision electroweak
data for constraining the Type~II 2HDM with a light Higgs boson.  
This data is known to provide an excellent fit to the Standard Model
with one Higgs doublet and $\mhl=86^{+48}_{-32}$~GeV \cite{erler}.
Nevertheless,
\Ref{Chankowski:1999ta} demonstrates that even with a light
Higgs mass below 20~GeV, the CP-conserving Type~II
2HDM provides an equally good fit to
the precision electroweak data.

One byproduct of this analysis is the potential for an improved
exclusion limit on the CP-odd Higgs boson mass in the region of light
$\mha$.  In the $\mha$ {\it vs.} $\tanb$ plane, the 
experimentally excluded region in a general 2HDM is essentially the
same as the shaded regions of fig.~1, based on the absence of 
$e^+e^-\to \ha f\bar f$ ($f=b$ or $\tau$) and $\Upsilon\to\ha\gamma$.  
If $\mha\ll\mhl$, $\mhh$, $\mhpm$, then \eq{app1} is replaced by:
\beq
\delta a_{\mu}^{\rm Higgs}\simeq 
\delta a_\mu^{{\ha}} \simeq \frac{-G_Fm_\mu^2}{4\pi^2 \sqrt{2}}
\left(\frac{m_\mu^2}{\mha^2}\right) \tan^2\beta
\bigg[
\ln\left(\frac{\mha^2}{m_\mu^2}\right)-\frac{11}{6} \biggr ] \;.
\label{app1a}
\eeq
The $\delta a_{\mu}^{\rm Higgs}$ contours shown in fig.~1 would apply
in this case [independent of the value of $\sinbma$] if each number
accompanying the contours is multiplied by $-0.9$ (approximately). 
Technically, one cannot use this to exclude
any region of $\mha$ {\it vs.} $\tanb$ parameter space, since the
negative contribution of \eq{app1a}
can be canceled by some positive contribution
(which by the assumption of \eq{newphys} must exist).  However, if a
future measurement were to establish that $\amu\simeq 0$, then barring
an accidental cancellation from more than one source of new physics, 
it would be possible to significantly extend the present excluded region
in the $\mha$ {\it vs.} $\tanb$ plane.

\section{Final Results and Conclusions}

If we combine the experimental bounds on the Higgs mass discussed in
section 3, 
we conclude that a light Higgs boson
can be responsible for the observed
$2.6\sigma$ deviation of the BNL measurement of the muon 
anomalous magnetic moment at the 90\%~CL in the framework of a
two-Higgs-doublet model with Model~II Higgs-fermion Yukawa couplings 
only if the model parameters satisfy the
following requirements:
\beqa
&& m_{\Upsilon}\lsim \mhl \lsim 2 m_B \;, \nonumber \\
&& \sinbma\simeq 0 \;, \nonumber \\
&& 30\lsim\tanb\lsim 35 \;.
\label{result}
\eeqa
In addition, $\hh$, $\ha$ and $\hpm$ must be sufficiently heavy to
satisfy the LEP experimental constraints.  In the model specified above,
the SM Higgs mass bound applies to $\hh$ so that $\mhh\gsim
113.5$~GeV.  The constraint on $\mha$ is deduced from the absence of
$Z\to\hl\ha$ (either by direct observation or as inferred from the
measured width of the $Z$), which implies that
$\mha\gsim 80$~GeV.\footnote{With
further LEP analysis, it might be possible to push the limit on
$\mha$ higher.
The large $\tanb$ MSSM Higgs analysis implies that $\mhl+\mha\gsim
180$~GeV due to the non-observation of $e^+e^-\to\hl\ha$.  However,
this analysis, which searches for $\hl\ha$ via a four jet 
topology, is highly inefficient for a very light $\hl$ and is thus
not applicable to the present model.} 
Finally, in a general 2HDM, $\mhpm\gsim 78.7$~GeV \cite{tomjunk}.

One noteworthy consequence of $\mhl\sim 10$~GeV is
the possibility of mixing
between the $\hl$ and the $0^{++}$ $b\bar b$ bound states
$\chi\ls{b0}(1P)$ and $\chi\ls{b0}(2P)$, as
discussed in \Refs{haber}{egns}.  As a result, the decay
$\chi\ls{b0}\to\tau^+\tau^-$ should be prominent.  The predicted
rate is roughly
\beq
{\Gamma(\chi\ls{b0}\to\tau^+\tau^-)\over\Gamma(\chi\ls{b0}\to~{\rm hadrons})}
\simeq {2.5\times 10^{-7}~{\rm GeV}^2\over
(m_\chi-\mhl)^2}\tan^4\beta\:,
\eeq
which is valid for $\mhl$ near $m_\chi$ but separated by a few Higgs
widths.\footnote{%
If the two masses are within a Higgs width, then the mixing
of the two states will be close to maximal \cite{egns}, 
and the corresponding $\tau^+\tau^-$ branching ratio of both 
eigenstates would be close to
100\% due to the large $\tan^4\beta$ enhancement.}
Due to the large $\tan^4\beta$ enhancement, the predicted branching ratio for 
$\chi\ls{b0}\to\tau^+\tau^-$ can be substantial.  Remarkably, the
Particle Data Group \cite{pdg} provides no data on possible 
decay modes of the $\chi\ls{b0}$ other than the radiative decays, 
$\chi\ls{b0}\to\Upsilon\gamma$, $\Upsilon'\gamma$.
 
Apart from a careful study of $\chi\ls{b0}$ decays, the 2HDM specified
by \eq{result} could be confirmed or ruled out by a more complete
analysis by the LEP Collaborations of their data in search of
$e^+e^-\to \hl f\bar f$ ($f=b$ or $\tau$).  We note that the ALEPH and
DELPHI exclusion plots used in fig.~1 are based on a preliminary
analyses and have not formally appeared in the literature.  Without
employing these LEP limits, the allowed 2HDM parameter space in which
$\hl$ contributes significantly to $\amu$ is substantially larger.  As
advocated in \Ref{kk}, the $\tanb$ exclusion limit could be lowered if
a complete analysis were performed using all of the LEP data.  The
potential significance of such a result should be clear from fig.~1.

In the absence of additional information from the LEP collider, one must
wait for a further improvement of the BNL measurement of the muon 
anomalous magnetic moment.  A factor of four increase in data is
expected when the data sets from the 2000 and 2001 runs are fully
analyzed .
If the significance of a nonzero result for $\amu$ increases, it will
be crucial to discover the source of the new physics.
To further constrain the Higgs sector contribution to $\amu$,
a high energy 
$e^+e^-$ linear collider that can perform precision studies of 
Higgs processes is required~\cite{Krawczyk:2000kf}.
One must either discover a light
Higgs boson with $\mhl\sim 10$~GeV or improve the 
present constraints in the $\mhl$ {\it vs.} $\tanb$ plane.

\vspace*{1cm}
\noindent {\bf \large Acknowledgments}
\vspace{0.5cm}

{\it We gratefully acknowledge Patrick Janot and Michael Kobel 
for useful discussions
concerning the LEP Higgs search.  We also thank Herbi Dreiner for 
his careful reading of the manuscript and a number of useful suggestions.
A.D. would like to acknowledge financial support from the
Network RTN European Program HPRN-CT-2000-0014
``Physics Across the Present Energy Frontier: Probing the Origin of
Mass.''  H.E.H. is supported in part by a grant from the
U.S. Department of Energy under contract DE-FG03-92ER40689.  Finally,
H.E.H. would like to thank H.P. Nilles and H.K. Dreiner for
their hospitality during his visit to the Physikalisches Institut der
Universit\"at Bonn, where this work was done.  
}

\end{document}